\documentclass[sigconf]{acmart}
\usepackage{algorithm}
\usepackage{algorithmic}
\usepackage{balance}

\AtBeginDocument{
  }

\copyrightyear{2026}
\acmYear{2026}
\setcopyright{cc}
\setcctype{by}
\acmConference[CIKM '26] {Proceedings of the 35th ACM International Conference on Information and Knowledge Management}{November 7--11, 2026}{Rome, Italy.}
\acmBooktitle{Proceedings of the 35th ACM International Conference on Information and Knowledge Management (CIKM '26), November 7--11, 2026, Rome, Italy}
\acmISBN{979-8-4007-2539-5/2026/11}
\acmDOI{10.1145/3799682.3840110}
\acmSubmissionID{924}
\begin{document}

\title{Efficient Generative Retrieval for E-commerce Search with Semantic Cluster IDs and Expert-Guided RL}

\author{Jianbo Zhu}
\authornote{Corresponding author.}
\affiliation{
  \institution{Taobao \& Tmall Group of Alibaba}
  \city{Hangzhou}
  \country{China}
}
\email{zhujianbo.zjb@taobao.com}

\author{Xing Fang}
\affiliation{
  \institution{Taobao \& Tmall Group of Alibaba}
  \city{Hangzhou}
  \country{China}
}
\email{fangxing.fx@taobao.com}

\author{Jing Wang}
\affiliation{
  \institution{Taobao \& Tmall Group of Alibaba}
  \city{Hangzhou}
  \country{China}
}
\email{jing.wangj1@taobao.com}

\author{Mingmin Jin}
\affiliation{
  \institution{Taobao \& Tmall Group of Alibaba}
  \city{Hangzhou}
  \country{China}
}
\email{jimmy.jmm@taobao.com}

\author{Bokang Wang}
\affiliation{
  \institution{Taobao \& Tmall Group of Alibaba}
  \city{Hangzhou}
  \country{China}
}
\email{wangbokang.wbk@taobao.com}

\author{Guangxin Song}
\affiliation{
  \institution{Taobao \& Tmall Group of Alibaba}
  \city{Hangzhou}
  \country{China}
}
\email{guangxin.sgx@taobao.com}

\author{Zhenyu Xie}
\affiliation{
  \institution{Taobao \& Tmall Group of Alibaba}
  \city{Hangzhou}
  \country{China}
}
\email{xuanyu.xzy@taobao.com}

\author{Junjie Bai}
\affiliation{
  \institution{Taobao \& Tmall Group of Alibaba}
  \city{Hangzhou}
  \country{China}
}
\email{baijunjie.bjj@taobao.com}

\renewcommand{\shortauthors}{Jianbo Zhu et al.}

\begin{abstract}
Generative retrieval offers a promising alternative by unifying the multi-stage retrieval process into a single end-to-end model. However, its practical adoption in industrial e-commerce search remains challenging, given the massive and dynamic product catalogs, strict latency requirements, and the need to align retrieval with downstream ranking goals. In this work, we propose a retrieval framework tailored for real-world recall scenarios, positioning generative retrieval as a recall-stage supplement rather than an end-to-end replacement. Our method, \textbf{CQ-SID} (Category-and-Query constrained Semantic ID), employs category-aware and query-item contrastive learning along with Residual Quantized VAEs to encode items into hierarchical semantic cluster identifiers, significantly improving click and exposure metrics and reducing beam search size and inference time. Additionally, we develop \textbf{EG-GRPO} (Expert-Guided Group Relative Policy Optimization), a reinforcement learning approach that aligns generative recall with downstream ranking under sparse rewards by injecting ground-truth samples to stabilize training. Offline experiments on TmallAPP search logs show that CQ-SID achieves up to 26.76\% and 11.11\% relative gains in semantic and personalized click hitrate over RQ-VAE baselines. EG-GRPO further improves multi-objective performance. Online A/B tests confirm gains in GMV (+1.15\%) and UCTCVR (+0.40\%). The generative recall channel now contributes substantially in production, accounting for 50.25\% of exposures, 58.96\% of clicks, and 72.63\% of purchases, demonstrating a viable path for deploying generative retrieval in real-world e-commerce systems.
\end{abstract}

%%
%% The code below is generated by the tool at http://dl.acm.org/ccs.cfm.
%% Please copy and paste the code instead of the example below.
%%


% \received{20 February 2007}
% \received[revised]{12 March 2009}
% \received[accepted]{5 June 2009}

\maketitle

\section{Introduction}
The architecture of modern e-commerce search engines typically follows a multi-stage funnel paradigm, comprising query understanding, recall, coarse ranking, fine ranking, and re-ranking stages. In this pipeline, the recall stage is responsible for efficiently retrieving a candidate set of potentially relevant items from a corpus of hundreds of millions of products. Traditional recall methods rely on sparse retrieval (e.g., BM25, inverted indices) and dense retrieval (e.g., approximate nearest neighbor search over embedding spaces), which achieve considerable success yet suffer from inherent limitations: sparse methods struggle with semantic matching, while dense methods require expensive index maintenance and suffer from the embedding gap between training and inference distributions.

Recently, generative retrieval has attracted significant attention as a new paradigm that replaces traditional index-then-retrieve architectures with a single differentiable model~\cite{tay2022transformer, wang2022neural, bevilacqua2022autoregressive}. In this paradigm, a sequence-to-sequence model directly generates document or item identifiers given a query, effectively collapsing indexing and retrieval into a unified generative process. The seminal work by Tay et al.~\cite{tay2022transformer} introduced the Differentiable Search Index (DSI), which demonstrated that transformer models could memorize and retrieve document identifiers with surprising efficacy. Subsequent works such as NCI~\cite{wang2022neural}, SEAL~\cite{bevilacqua2022autoregressive}, and TIGER~\cite{rajput2023recommender} have extended this paradigm to various domains and identifier designs.

Despite academic progress, deploying generative retrieval in industrial e-commerce search entails unique challenges that remain under-explored:

\textbf{Challenge 1: End-to-end feasibility at scale.} Prior industrial approaches such as the Kuaishou OneModel series~\cite{deng2025onerec, zhou2025onerec, chen2025onesearch} advocate end-to-end generation that bypasses the multi-stage funnel entirely. While this is viable in recommendation scenarios where exploration and diversity are inherently valued, search demands high precision and relevance. Generating directly from a corpus of hundreds of millions of items raises serious concerns about coverage, latency, and the ``Matthew effect'' (over-concentration on popular items).

\textbf{Challenge 2: Semantic ID design trade-offs.} Existing methods typically pursue a one-item-one-ID mapping to minimize collision rates~\cite{tay2022transformer, rajput2023recommender}. Yet in industrial settings, overly fine-grained identifiers require substantially more training data and longer convergence times. During inference, achieving sufficient recall breadth necessitates large beam sizes, creating latency and resource bottlenecks that are incompatible with production constraints.

\textbf{Challenge 3: Alignment with downstream ranking.} Recall models have traditionally been trained to optimize click-through signals via maximum likelihood estimation. However, a good recall model should not only retrieve clicked items but also expose high-quality items that downstream ranking can leverage. Bridging the optimization gap between recall and ranking remains an open problem.

To address these challenges, we position generative retrieval as a recall-stage supplement rather than an end-to-end replacement. Our key contributions are:

\begin{itemize}
\item We propose \textbf{CQ-SID}, a Category-and-Query constrained Semantic ID constructed via RQ-VAE with category-aware residual quantization and Query-Item contrastive learning. Unlike prior work that pursues unique IDs, CQ-SID treats each ID as a semantic cluster identifier, balancing discriminability with aggregation.
\item We design a \textbf{progressive training pipeline} that gradually builds query-to-SID mapping capabilities through supervised fine-tuning on item titles, query-item associations, and user personalization features, using a lightweight Qwen2.5-0.5B~\cite{qwen2025qwen25technicalreport} backbone.
\item We introduce \textbf{EG-GRPO}, an Expert-Guided Group Relative Policy Optimization method that aligns generative recall with downstream ranking signals. By injecting ground-truth samples into the policy gradient group, EG-GRPO stabilizes learning in sparse-reward environments and improves both hitrate and pvr.
\item We present extensive offline experiments and large-scale online A/B tests on a major mobile e-commerce platform, demonstrating that the proposed framework achieves superior recall quality with reduced computational cost and significant business metric improvements.
\end{itemize}

\begin{figure*}[t]
\includegraphics[width=0.65\textwidth]{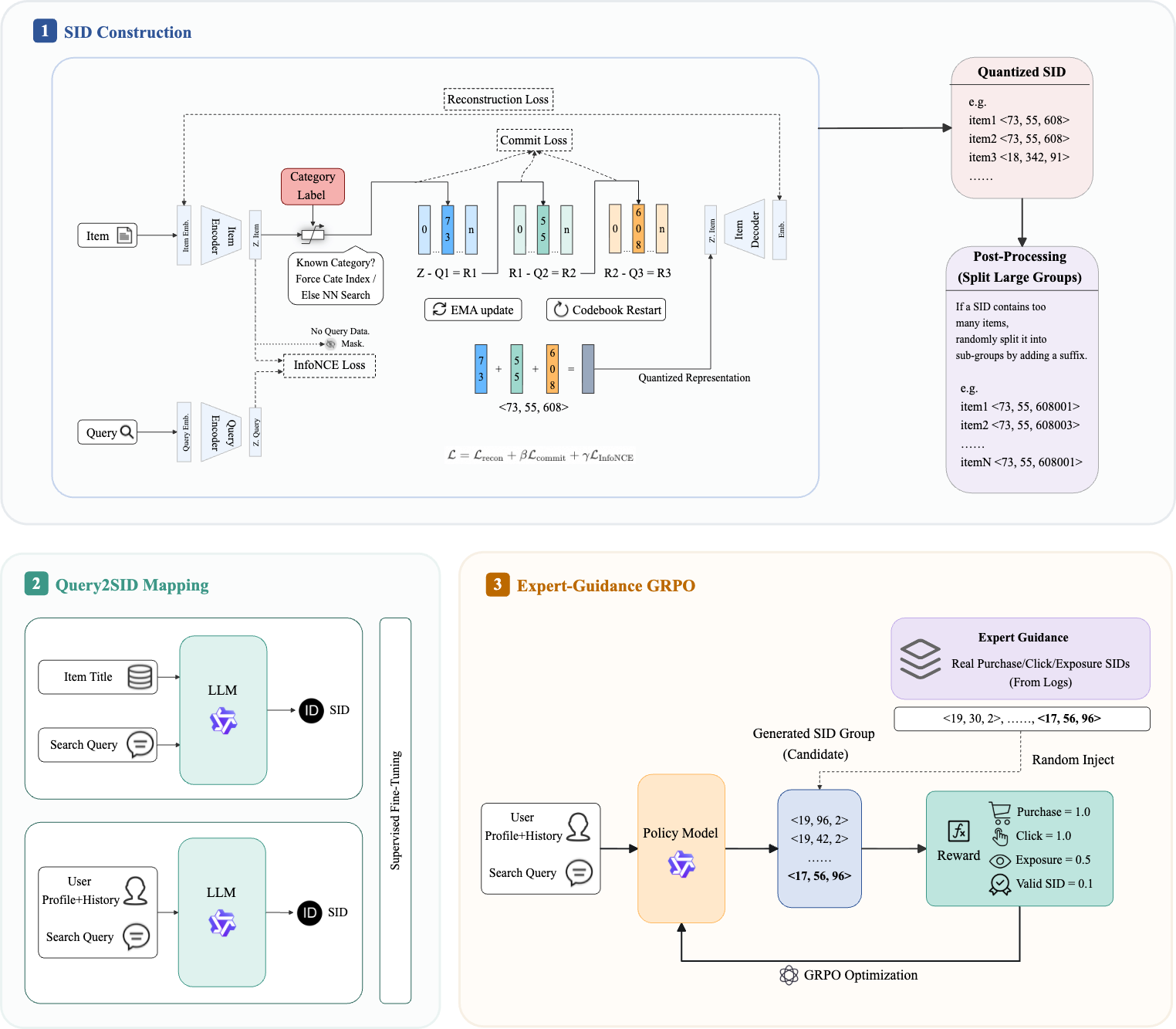}
\caption{Architecture overview of our generative recall framework, showing the pipeline: (1) CQ-SID item encoding, (2) progressive query-to-SID learning, and (3) EG-GRPO ranking alignment.}
\label{fig:framework}
\end{figure*}

\section{Related Work}

\subsection{Generative Retrieval}

The field of generative information retrieval has evolved rapidly since the introduction of DSI by Tay et al.~\cite{tay2022transformer}, which proposed to train a single transformer model to map queries directly to document identifiers, effectively replacing the traditional indexing-retrieval pipeline.  Sun et al.~\cite{sun2023learning} investigated learned tokenization schemes for generative retrieval, while Li et al.~\cite{li2024survey} and Kuo et al.~\cite{kuo2024survey} provided comprehensive surveys of the rapidly expanding GenIR landscape. Recent work has further revealed theoretical connections between generative retrieval and multi-vector dense retrieval~\cite{wu2024generative}.

The application of generative retrieval to e-commerce search and recommendation has gained substantial industrial interest. Kuaishou's OneModel frameworks~\cite{deng2025onerec, zhou2025onerec, chen2025onesearch} represent ambitious attempts to unify the entire search and recommendation pipeline under a single generative model. In the broader recommendation literature, structured multi-behavior modeling and dynamically evolving user preference signals have likewise been explored~\cite{zhang2025hierarchical, cui2023dynamic}. Beyond end-to-end unification, recent work has also examined the integration of generative retrieval with structured taxonomies in e-commerce. Yang et al.~\cite{yang2025gsid} proposed GSID, a generative semantic indexing framework that leverages product attribute hierarchies to guide identifier learning for e-commerce product understanding. Meanwhile, Liu et al.~\cite{liu2026cat} introduced CAT-ID$^2$, which explicitly encodes category trees into document identifiers to improve semantic structure and retrieval accuracy in e-commerce generative retrieval.

Nevertheless, existing works predominantly treat generative retrieval as a standalone replacement for conventional pipelines, leaving open the question of how such models can coexist with and complement industrial multi-stage funnel architectures under stringent latency and coverage constraints. 

\subsection{Semantic Identifiers}

A critical design choice in generative retrieval is the construction of document or item identifiers. Early approaches used arbitrary numeric strings or titles as identifiers~\cite{tay2022transformer, bevilacqua2022autoregressive}, which suffered from poor generalization and high collision rates. The introduction of semantic identifiers based on vector quantization marked a significant advancement. Rajput et al.~\cite{rajput2023recommender} proposed TIGER, which employed Residual-Quantized Variational AutoEncoders (RQ-VAE) to learn semantically meaningful, hierarchical item identifiers for recommendation. This approach was subsequently extended by various works~\cite{dai2025rq, fu2025forge} exploring improved quantization strategies and identifier structures.

Several recent works have specifically addressed e-commerce scenarios. Wu et al.~\cite{wu2024hi} proposed Hi-Gen, a hierarchical encoding-decoding generative retrieval method for large-scale personalized e-commerce search. Fu et al.~\cite{fu2025forge} introduced FORGE, utilizing i2i collaborative contrast enhancement and collision handling to enhance SID quality. Several works have further advanced identifier design by jointly optimizing for discriminability and semantic expressiveness. Cheng et al.~\cite{cheng2025descriptive} proposed D$^2$-DocID, which constructs both descriptive and discriminative document identifiers to balance human interpretability and model distinguishability in generative retrieval.

Despite these advances, the tension between identifier granularity, which dictates collision rates and semantic expressiveness, and the resulting inference cost remains insufficiently explored, particularly in settings where identifiers must accommodate dynamically evolving item corpora at scale.

\subsection{Reinforcement Learning}

Aligning retrieval models with downstream objectives has been extensively studied through the lens of reinforcement learning and preference optimization. The classical reinforce algorithm~\cite{williams1992simple} and its successor PPO~\cite{schulman2017proximal} have been applied to various information retrieval tasks. More recently, Direct Preference Optimization (DPO)~\cite{rafailov2023direct} and Group Relative Policy Optimization (GRPO)~\cite{shao2024deepseekmath} have emerged as efficient alternatives that eliminate the need for explicit value networks. GRPO, in particular, computes advantages through group-based normalization, making it especially suitable for scenarios with sparse rewards.

However, applying these methods to generative retrieval remains nontrivial, as the discrete and structured output space of semantic identifiers introduces unique challenges in reward design and policy gradient variance that standard RL formulations do not directly address.

\section{Methodology}

Our generative recall system comprises three core stages: (1) item semantic ID construction via CQ-SID, (2) progressive user-personalized query-to-SID mapping, and (3) ranking-aligned reinforcement learning via EG-GRPO. Figure~\ref{fig:framework} provides an architectural overview of the proposed framework.

\subsection{Constructing Item Semantic IDs via CQ-SID}
\subsubsection{Motivation and Design Principles}

Rather than pursuing unique item identifiers as in prior work~\cite{tay2022transformer, rajput2023recommender}, we design CQ-SID (Category-and-Query constrained Semantic ID) to achieve three key properties:
\begin{itemize}
\item \textbf{Semantic aggregation:} Semantically or attribute-similar items are mapped to a shared SID, subject to a predefined cap on the number of items per SID.
\item \textbf{Inter-cluster discriminability:} Distinct semantic items should map to distinct SIDs, ensuring sufficient granularity.
\item \textbf{Hierarchical structure:} Items with shared characteristics should share SID prefixes, thereby establishing an efficient semantic hierarchical structure.
\end{itemize}

This cluster-based design relaxes the collision-free requirement to substantially reduce beam search sizes and improve coverage, a trade-off we argue is essential for industrial-scale deployment.

\subsubsection{Model Architecture}
We build CQ-SID upon RQ-VAE~\cite{zeghidour2021soundstream, rajput2023recommender} with three key innovations.
\paragraph{(1) Category-Guided Residual Quantization.}
In e-commerce scenarios, the item category system offers a natural structural basis for hierarchical partitioning of SIDs. Therefore, the first-level codebook is designed to be category-aware, where top-level categories serve as quantization constraints at the first stage, resulting in 1,711 valid categories. Overall, we adopt a three-level residual vector quantization framework with codebook sizes $K_1 \times K_2 \times K_3 = 2048 \times 1024 \times 1024$.

During training of the first-level quantization, for items with known categories, their category labels are mandatorily used as indexes for updating the codebook vectors:
\begin{equation}
k_{i}^{(1)} = \begin{cases}
\text{CategoryID}(i), & \text{if } i \in \mathcal{I}_{\text{known}}, \\
\underset{j \in \{1, \ldots, K_1\}}{\arg\min} \|\mathbf{r}_i^{(0)} - \mathbf{e}_j^{(1)}\|_2^2, & \text{otherwise},
\end{cases}
\end{equation}

where $\mathbf{r}_i^{(0)}$ is the item encoder embedding, $\mathbf{e}_j$ is the first-level codebook vector.

The remaining second and third level codebooks adopt standard nearest-neighbor quantization. All codebook parameters are updated using Exponential Moving Average (EMA). To avoid codebook collapse, we employ a codebook restart strategy~\cite{van2017neural}.
For the code vector $\mathbf{e}_j^{(l)}$ in the $l$-th level codebook, its EMA update formula is:

\begin{align}
\mathbf{e}_j^{(l)} &\leftarrow \lambda \, \mathbf{e}_j^{(l)} + (1 - \lambda) \, \mathbf{m}_j^{(l)}, \\
\mathbf{m}_j^{(l)} &= \frac{1}{n_j^{(l)}} \sum_{i \in \mathcal{B}} \mathbb{I}\big[k_i^{(l)}=j\big] \, \mathbf{r}_{i}^{(l-1)},
\end{align}

where $\mathbf{r}_{i}^{(l-1)}$ and $k_i^{(l)}$ represent the residual vector and index for the $i$-th sample at the $l$-th quantization level, $n_j^{(l)} = \sum_{i \in \mathcal{B}} \mathbb{I}[k_i^{(l)}=j]$ is the number of samples allocated to index j in the current batch, and $\lambda \in (0,1)$ is the EMA decay factor.

\paragraph{(2) Query-Item Contrastive Learning.}

The CQ-SID model simultaneously receives embeddings of both items and queries, utilizing high-confidence query-item pairs (derived from user search, click, and purchase behaviors) to enhance semantic alignment through bidirectional InfoNCE loss~\cite{oord2018representation}. This enables similar items to become more closely associated via their shared relationships with queries.
\begin{align}
\mathcal{L}_{\text{Bi-InfoNCE}} = \frac{1}{2} \bigg[ 
 & -\log \frac{\exp(\text{sim}(\mathbf{e}_i, \mathbf{e}_q) / \tau)}{\sum_{q' \in \mathcal{Q}} \exp(\text{sim}(\mathbf{e}_i, \mathbf{e}_{q'}) / \tau)} \nonumber \\
 & -\log \frac{\exp(\text{sim}(\mathbf{e}_q, \mathbf{e}_i) / \tau)}{\sum_{i' \in \mathcal{I}} \exp(\text{sim}(\mathbf{e}_q, \mathbf{e}_{i'}) / \tau)} \bigg],
\end{align}

where $\mathcal{Q}$ and $\mathcal{I}$ denote the in-batch query and item sample sets, $\text{sim}(\cdot, \cdot)$ is cosine similarity, and $\tau$ is the temperature coefficient. For items without associated queries, the contrastive term is masked to maintain training stability.

\paragraph{(3) Joint Training Objective.}

The total loss is a weighted sum of reconstruction, commitment, and contrastive losses:

\begin{gather}
\mathcal{L}_{\text{recon}} = \|x - \hat{x}\|_2^2, \\
\mathcal{L}_{\text{commit}} = \sum_{i=1}^3 \|\mathbf{z}_i - \text{sg}[\hat{\mathbf{z}}_i]\|_2^2, \\
\mathcal{L} = \mathcal{L}_{\text{recon}} + \beta \mathcal{L}_{\text{commit}} + \gamma \mathcal{L}_{\text{InfoNCE}},
\end{gather}

where $x$ is the input feature, $\hat{x}$ the reconstruction, $\mathbf{z}_i$ the quantizer input, $\hat{\mathbf{z}}_i$ the quantized representation, $\text{sg}[\cdot]$ the stop-gradient operator, and $\beta, \gamma$ are hyperparameters.

\subsubsection{SID Post-Processing}
To control cluster size, we apply post-processing: if the number of items for a SID exceeds threshold $T_{\max}$, the third-level indices are randomly grouped. Given $c$ items under SID $\langle s_1, s_2, s_3 \rangle$, the number of groups is 
% \[
% G = \min\big(\lceil c / T_{\max} \rceil, G_{\max}\big)
% \].
$G = \min\big(\lceil c / T_{\max} \rceil, G_{\max}\big)$. Items are randomly partitioned into $G$ sub-clusters, and a group suffix is concatenated. This preserves the hierarchical prefix structure while preventing single popular clusters from dominating retrieval.

\textbf{Example:} Original SID $\langle s_1^{(73)}, s_2^{(55)}, s_3^{(608)} \rangle$ \\
$\rightarrow$ New SIDs: $\langle s_1^{(73)}, s_2^{(55)}, s_3^{(0608001)} \rangle$, $\langle s_1^{(73)}, s_2^{(55)}, s_3^{(0608002)} \rangle$, \ldots

\begin{algorithm}
\caption{Post-processing Strategy for SID Grouping}
\label{alg:postprocessing}
\begin{algorithmic}
\REQUIRE Set of SIDs $S$, threshold $T_{\max}=50$, maximum groups $G_{\max}=100$
\FOR{each SID $s = \langle s_1, s_2, s_3 \rangle \in S$}
\STATE $c \gets$ number of items associated with $s$
\IF{$c > T_{\max}$}
\STATE $G \gets \min\left(\left\lceil \frac{c}{T_{\max}} \right\rceil, G_{\max}\right)$
\STATE Randomly partition items into $G$ subgroups
\FOR{$g = 1$ to $G$}
\STATE $s_3' \gets \text{format}(s_3, \text{`004d'}) \oplus \text{format}(g, \text{`003d'})$ \COMMENT{4-digit base + 3-digit group ID}
\STATE Create new SID: $\langle s_1, s_2, s_3' \rangle$
\ENDFOR
\ENDIF
\ENDFOR
\end{algorithmic}
\end{algorithm}

\textbf{Why random partitioning is reasonable.} After CQ-SID training, items under the same SID already share highly similar semantic and category features. Consequently, splitting a large cluster through random post-processing does not excessively break semantic coherence. This step mainly prevents any single SID from accumulating too many items, which would otherwise skew the downstream model’s beam search toward a narrow semantic range and heighten the risk of the Matthew effect in retrieval.

\subsection{Progressive Query-to-SID Learning}

We train the query-to-SID mapping model in four progressive stages using Qwen2.5-0.5B, a lightweight yet capable language model.
\paragraph{Stage 1: Item-to-SID Mapping.}
Using the mapping $\phi$ from CQ-SID, we construct (item title, SID) pairs and perform supervised fine-tuning (SFT). This stage establishes the basic understanding of the relationship between item text descriptions and semantic identifiers.

\paragraph{Stage 2: Query-to-SID Mapping.}
We leverage query-item association data to construct (query, SID) pairs. For each query, we randomly sample $N=3$ SIDs from its associated clicked/purchased items as target outputs. This stage teaches the model to map search intent to product semantic clusters.

\paragraph{Stage 3: Personalized (User+Query)-to-SID Mapping.}
To support personalized recall, we augment the query with user features (gender, age group) and the SID sequence of recently clicked items under related categories. This enables the model to generate user-specific SID lists for the same query.

\paragraph{Stage 4: Ranking-Aligned Refinement via EG-GRPO.}
While Stages 1-3 optimize the log-likelihood via supervised fine-tuning for click and purchase signals, recall models must also expose diverse, high-quality candidates for downstream ranking. We therefore apply reinforcement learning to align the generative model with ranking-oriented objectives, as detailed in Section~\ref{sec:eg-grpo}.

\subsection{Expert-Guided GRPO}
\label{sec:eg-grpo}

We adopt Group Relative Policy Optimization (GRPO)~\cite{shao2024deepseekmath} to align recall output with ranking signals. For input $\mathbf{x}$, the model generates a group of outputs $\{o_1, \dots, o_G\}$. Each output receives a reward based on its relationship to ground-truth user behavior:
\begin{equation}
R(o) = \begin{cases}
1.0, & \text{if } o \in \mathcal{P}_{\text{pay}}(x), \\
1.0, & \text{else if } o \in \mathcal{P}_{\text{clk}}(x), \\
0.5, & \text{else if } o \in \mathcal{P}_{\text{exp}}(x), \\
0.1, & \text{else if } o \in \mathcal{S}_{\text{valid}}, \\
0.0, & \text{otherwise},
\end{cases}
\end{equation}

where $\mathcal{P}_{\text{pay}}$, $\mathcal{P}_{\text{clk}}$, $\mathcal{P}_{\text{exp}}$ denote purchased, clicked, and exposed item SID sets, and $\mathcal{S}_{\text{valid}}$ is the valid SID set.

The advantage function is computed via group normalization:
\begin{equation}
A(o_i) = \frac{R(o_i) - \text{mean}(\{R(o_j)\}_{j=1}^G)}{\text{std}(\{R(o_j)\}_{j=1}^G) + \epsilon}.
\end{equation}

The standard GRPO objective follows clipped policy gradient:
\begin{align}
\mathcal{L}_{\text{GRPO}}(\theta) = & -\frac{1}{G} \sum_{i=1}^G \min\left(\frac{P_\theta(o_i \mid x)}{P_{\theta_{\text{old}}}(o_i \mid x)} A(o_i), \right. \nonumber \\
& \left. \text{clip}\left(\frac{P_\theta(o_i \mid x)}{P_{\theta_{\text{old}}}(o_i \mid x)}, 1-\epsilon_{\text{clip}}, 1+\epsilon_{\text{clip}}\right) A(o_i)\right).
\end{align}

Because click and purchase signals are extremely sparse in search logs, standard GRPO often suffers from high variance and unstable optimization. We therefore propose \textbf{EG-GRPO}: in each group, we randomly sample $K$ ground-truth SIDs from $\mathcal{P}_{\text{clk}} \cup \mathcal{P}_{\text{exp}}$ and inject them as pseudo-generated outputs. The expanded group $\mathcal{G} = \mathcal{G}_{\text{sampled}} \cup \mathcal{G}_{\text{expert}}$ participates in reward computation and gradient updates, providing stable, high-quality optimization signals. This expert injection serves as a form of implicit curriculum learning, ensuring that each policy gradient update incorporates positive examples even when the model's own samples are poor.

\begin{algorithm}
\caption{Expert-Guided GRPO (EG-GRPO)}
\label{alg:eg_grpo}
\begin{algorithmic}
\REQUIRE Pre-trained model $M$, training dataset $\mathcal{D}_{\text{RL}}$, group size $G$, number of expert samples $K$
\STATE Initialize policy model $\pi_\theta \gets M$
\FOR{each epoch}
    \FOR{each batch $\{x_1, \ldots, x_B\} \subset \mathcal{D}_{\text{RL}}$}
        \FOR{$b = 1$ to $B$}
            \STATE Sample $G$ outputs: $\mathcal{G}_{\text{sampled}}^{(b)} \sim \pi_{\theta_{\text{old}}}$
            \STATE Sample $K$ expert SIDs: $\mathcal{G}_{\text{expert}}^{(b)} \sim \mathcal{P}_{\text{clk}}(x_b) \cup \mathcal{P}_{\text{exp}}(x_b)$
            \STATE $\mathcal{G}^{(b)} \gets \mathcal{G}_{\text{sampled}}^{(b)} \cup \mathcal{G}_{\text{expert}}^{(b)}$
            \STATE Compute rewards $R(o)$ for $o \in \mathcal{G}^{(b)}$
            \STATE Compute advantages $A(o)$ via group normalization
        \ENDFOR
        \STATE Compute GRPO loss $\mathcal{L}_{\text{GRPO}}$
        \STATE Update $\theta$ using $\nabla_\theta \mathcal{L}_{\text{GRPO}}$
    \ENDFOR
\ENDFOR
\end{algorithmic}
\end{algorithm}

\subsection{Online Inference and Items Filtering}

During online serving, user query and profile features are assembled, and the model performs beam search to generate top-$K$ candidate SIDs. These SIDs are mapped to actual items via a pre-built SID-to-Items lookup table.

To ensure quality, we filter the full item corpus (hundreds of millions) to a high-efficiency subset of approximately 21 million items as the generative recall pool. Within this pool, the resulting SID-to-items mapping contains an average of 4.2 items per SID. This pool is dynamically updated daily: newly efficient items are inferenced through CQ-SID and attached to their semantic clusters.

\section{Experiments}

\begin{table*}[t]
\centering
\caption{Click hitrate of semantic retrieval at same beam size.}
\label{tab:semantic_beam}
\begin{tabular}{lccc}
\toprule
\textbf{Method} & \textbf{beam@1} & \textbf{beam@10} & \textbf{beam@100} \\
\midrule
RQ-VAE & 0.0598 & 0.2579 & 0.5199 \\
CQ-SID (w/o Cate) & 0.0680 (+13.71\%) & 0.2870 (+11.28\%) & 0.5578 (+7.29\%) \\
CQ-SID (w/o QI) & 0.0596 (-0.33\%) & 0.2691 (+4.34\%) & 0.5652 (+8.71\%) \\
\textbf{CQ-SID} & \textbf{0.0758 (+26.76\%)} & \textbf{0.3161 (+22.57\%)} & \textbf{0.6181 (+18.89\%)} \\
\bottomrule
\end{tabular}
\end{table*}

\subsection{Experimental Setup}
\paragraph{Dataset.} All experiments were conducted on real search logs from a major mobile e-commerce platform. The CQ-SID model was trained on 37.5M samples (21.1M with query-item pairs, 16.4M item-only). The LLM progressive model was trained on 21.0M samples for Item2SID, 90.3M for Query2SID, and 73.7M for personalized UQ2SID. The test set contained 201k samples for Query2SID and 170k for UQ2SID evaluation.

\paragraph{Metrics.} We primarily report Hitrate. Since different SID schemes yield varying numbers of items per SID, we evaluate along two dimensions: (1) same beam size, measuring inference-efficiency-adjusted quality; and (2) top-1K items after efficiency-score truncation, simulating online truncation logic.

\paragraph{Baselines.} The primary baseline is the RQ-VAE~\cite{zeghidour2021soundstream} used in TIGER~\cite{rajput2023recommender}. We also perform ablation studies removing category constraints and query-item contrastive learning individually. All LLM-based SID mapping models use Qwen2.5-0.5B.

\paragraph{Implementation Details.} CQ-SID uses codebook sizes [2048, 1024, 1024], trained on 4 GPUs for 10 epochs with batch size 4096 and cosine learning rate $10^{-3}$. Loss weights are $\beta=1.0$, $\gamma=0.001$, temperature $\tau=0.1$, and contrastive sample size $b=128$. LLM models are trained on 64 GPUs: I2SID uses batch size 128, LR $10^{-4}$, 2k steps; Q2SID uses batch size 256, LR $4\times10^{-5}$, 2.5k steps; UQ2SID uses batch size 64, LR $4\times10^{-5}$, 5k steps. EG-GRPO uses rollout batch size 512, group size 8, KL weight 1.0, LR $10^{-6}$, 1k steps.

\subsection{Semantic Generative Recall Results}
\subsubsection{Same Beam Size Comparison}
Beam search size directly determines computational resource consumption during inference. We compared the hitrate of RQ-VAE and CQ-SID under identical beam size conditions, with ablation studies isolating the effects of category constraints and query-item contrastive learning.

Table~\ref{tab:semantic_beam} demonstrates that CQ-SID consistently outperforms the standard RQ-VAE baseline across all beam sizes. The most pronounced gain occurs at beam@1, where CQ-SID achieves a \textbf{26.76\%} relative improvement in hitrate. This indicates that the combination of category constraints and query-item contrastive learning substantially enhances the semantic quality of SIDs, enabling the model to identify the correct semantic cluster even with minimal decoding budget.
The ablation study reveals complementary effects of the two design components. Removing category constraints (CQ-SID w/o Cate) still yields substantial gains (+13.71\% at beam@1), confirming that query-item contrastive learning alone provides meaningful semantic alignment. Conversely, removing contrastive learning (CQ-SID w/o QI) preserves moderate improvements under larger beam sizes, attributable to the hierarchical structure induced by category-aware quantization. The synergy between both modules, semantic clustering through categories and cross-modal alignment through contrastive learning, drives the full performance gains.

\subsubsection{Top-1K Truncation Comparison}
Because different SID schemes associate varying numbers of items with each identifier, comparing hitrate under identical beam sizes may conflate identifier quality with candidate set size. To disentangle these effects, we generate varying numbers of beams for each query, truncate candidates based on product efficiency scores, and retain the top-1K items for comparison. Table~\ref{tab:semantic_top1k} reports the results.

\begin{table}[!htb]
\centering
\caption{Click hitrate at top-1K truncation.}
\label{tab:semantic_top1k}
\begin{tabular}{lcccccc}
\toprule
\textbf{Method} & \textbf{b@25} & \textbf{b@30} & \textbf{b@35} & \textbf{b@60} & \textbf{b@65} & \textbf{b@70} \\
\midrule
RQ-VAE & 0.3675 & 0.3870 & 0.4016 & 0.4272 & \textbf{0.4275} & 0.4272 \\
\textbf{CQ-SID} & 0.4370 & \textbf{0.4422} & 0.4403 & 0.4001 & 0.3911 & 0.3825 \\
\bottomrule
\end{tabular}
\end{table}

Under the top-1K truncation setting, RQ-VAE attains its peak hitrate of 0.4275 at beam size 65, whereas CQ-SID reaches a superior hitrate of 0.4422 at beam size 30, representing a \textbf{3.44\%} relative improvement with \textbf{53.85\%} fewer beams. This efficiency gain is critical for production deployment: smaller beam sizes directly translate to lower inference latency and reduced computational cost without compromising recall quality.

The non-monotonic hitrate trends observed as beam size increases warrant careful interpretation. This phenomenon can be explained from two perspectives. On the one hand, in offline experiments, truncation is directly based on item efficiency, which differs from the actual online recall process that employs coarse-ranking truncation. On the other hand, this discrepancy leads to a critical issue: excessive low-confidence candidates dilute the candidate pool, and downstream ranking stages may lack sufficient discriminative capacity to filter them effectively, ultimately harming the overall system performance. This pattern validates the design philosophy that recall breadth must be balanced against candidate quality, particularly in funnel-based ranking architectures.

\begin{table*}[ht]
\caption{Click hitrate of personalized generative retrieval results.}
\label{tab:personalized}
\centering
\begin{tabular}{lcccc}
\toprule
\multicolumn{5}{c}{\textit{Same Beam Size Comparison}} \\
\midrule
Method & beam@1 & beam@10 & beam@50 & beam@100 \\
\midrule
RQ-VAE & 0.1359 & 0.4787 & 0.6912 & 0.7513 \\
\textbf{CQ-SID} & \textbf{0.1510 (+11.11\%)} & \textbf{0.5206 (+8.75\%)} & \textbf{0.7431 (+7.51\%)} & \textbf{0.8062 (+7.31\%)} \\
\midrule
\multicolumn{5}{c}{\textit{Top-1K Truncation Comparison}} \\
\midrule
Method & beam@155 & beam@160 & beam@190 & beam@195 \\
\midrule
RQ-VAE & 0.7567 & 0.7575 & 0.7604 & \textbf{0.7607} \\
\textbf{CQ-SID} & 0.7983 & \textbf{0.7984} & 0.7977 & 0.7975 \\
\bottomrule
\end{tabular}
\end{table*}

\begin{table*}[ht]
\centering
\caption{Effect of EG-GRPO on ranking alignment.}
\label{tab:rl}
\begin{tabular}{lccccc}
\toprule
\textbf{Method} & \textbf{clk@1} & \textbf{clk@10} & \textbf{exp@1} & \textbf{exp@10} & \textbf{pvr@10} \\
\midrule
CQ-SID & 0.1510 & 0.5206 & 0.5056 & 0.8693 & 0.4371 \\
+ GRPO (K=0) & 0.1519 & 0.5196 & 0.5077 & 0.8702 & 0.4360 \\
+ EG-GRPO (K=2) & \textbf{0.1524} & \textbf{0.5221} & \textbf{0.5091} & 0.8703 & 0.4377 \\
+ EG-GRPO (K=4) & 0.1523 & 0.5219 & 0.5087 & \textbf{0.8711} & \textbf{0.4378} \\
\bottomrule
\end{tabular}
\end{table*}

\subsection{Personalized Generative Recall Results}
Different users may exhibit distinct interests in response to the same query. While ranking stages can incorporate personalization, it is equally crucial for the recall stage to retrieve a diverse set of candidates to maximize overall system performance. To this end, we enhance query inputs with personalized signals—such as user profile and recent interaction histories from related categories.

Similar to the semantic generative recall above, we compared the hitrate of different methods under personalized scenarios with the same beam size and the same top1K truncation.

Table~\ref{tab:personalized} shows that incorporating personalization signals substantially improves hitrate for both methods, underscoring the value of user historical behavior for intent prediction. Under identical beam sizes, CQ-SID maintains stable advantages over RQ-VAE across all configurations, achieving relative improvements of 7\%--11\% from beam@1 to beam@100.

In the top-1K truncation comparison, RQ-VAE achieves its highest hitrate of 0.7607 at beam size 195, whereas CQ-SID attains 0.7984 at beam size 160. This represents a \textbf{4.96\%} relative hitrate improvement alongside a \textbf{17.95\%} reduction in beam size. These results further confirm that CQ-SID achieves a superior trade-off between computational efficiency and recall quality in personalized retrieval settings.

\subsection{Ranking Alignment via EG-GRPO}
The experiments are based on CQ-SID personalized generative recall. To verify the effectiveness of ranking-aligned refinement, we compare click hitrate (clk), exposure hitrate (exp), and the average exposure coverage ratio (pvr) under beam@10. Here, $K$ represents the number of ground-truth samples injected as expert guidance.

Table~\ref{tab:rl} reveals three important findings:
\begin{itemize}
\item Standard GRPO ($K=0$) shows an asymmetric pattern: slight gains in click hitrate at beam@1 but declines at beam@10, alongside a drop in pvr. This stems from a \emph{mode concentration} effect due to sparse rewards. Without expert guidance, high policy gradient variance causes the model to concentrate probability on a narrow set of high-confidence SIDs, sharpening top-1 predictions (improving clk@1 and exp@1). However, this reduces diversity at deeper beams, with similar candidates being repeated instead of exploring more semantic clusters, leading to degraded clk@10 and exposure coverage. Essentially, GRPO collapses into an exploitation-heavy regime, sacrificing the exploratory breadth needed for recall.

\item Introducing expert guidance ($K=2$ and $K=4$) yields consistent improvements in click hitrate, exposure hitrate, and exposure coverage. EG-GRPO leverages ground-truth samples as high-quality exemplars within the policy gradient group, effectively guiding the model to optimize precision-related objectives while preserving recall breadth. This validates our hypothesis that expert injection mitigates variance-induced collapse in sparse-reward settings.

\item Moderate expert guidance proves beneficial. As $K$ increases from 2 to 4, click-related metrics plateau. A seesaw effect emerges in exposure metrics between a small number of items (beam@1) and a larger number (beam@10). Nonetheless, both configurations improve over the non-expert baseline, indicating that even limited expert supervision is sufficient to stabilize policy optimization.
\end{itemize}

\textbf{On the magnitude of improvement.} While the absolute gains of EG-GRPO appear modest in Table~\ref{tab:rl}, the consistently positive deltas across all metrics indicate a qualitative behavioral shift: expert injection successfully redirects the optimization trajectory from exploitative collapse toward balanced exploration. The limited magnitude is structurally expected for three reasons. First, the three-stage SFT pipeline already brings the query-to-SID mapping close to a strong local optimum, leaving inherently narrow headroom for RL-based refinement. Second, hitrate is a binary metric: once the target SID appears in the beam, no further improvement is possible for that query. Third, EG-GRPO navigates the non-trivial trade-off between click and exposure, the collapse observed under standard GRPO ($K=0$) confirms that this trade-off is real and sharp. Achieving concurrent improvements across these metrics (a joint Pareto improvement) is substantially harder than maximizing any single metric in isolation. This capability represents EG-GRPO's primary methodological contribution.

\subsection{Online A/B Test}
We deployed the generative recall system in production search on TmallAPP, a major mobile e-commerce application. The system operates with dynamic beam sizes in [20, 50, 100], serving over 200 QPS across 8 GPUs with an online service availability of 99.9\% and an average end-to-end latency of approximately 40ms. Results from a two-week online A/B test comparing our method with and without CQ-SID confirmed the effectiveness of the proposed method in the multi-channel recall stage, with statistically significant improvements in key metrics: GMV +1.15\%, UCTCVR +0.40\%. Currently, the system is fully deployed in TmallAPP search product line, and among all recall channels, the generative recall channel accounts for 50.25\% of all exposures, 58.96\% of all clicks, and 72.63\% of all purchases under overlap-inclusive attribution.

\section{Conclusion}
This paper presents a generative retrieval framework, comprising CQ-SID and EG-GRPO, designed for the industrial e-commerce search recall stage. By treating semantic IDs as cluster identifiers rather than unique item labels, CQ-SID based on category constraints and contrastive learning achieves substantially higher recall quality with reduced inference cost compared to standard RQ-VAE baselines. Our progressive training pipeline enables effective query-to-SID mapping, while EG-GRPO aligns recall outputs with downstream ranking signals through expert-guided reinforcement learning, thereby modestly yet consistently enhancing multi-objective performance. Extensive offline experiments and large-scale online A/B tests demonstrate consistent improvements in hitrate, pvr, and business metrics, with the system currently deployed in TmallAPP production.

\section{GenAI Usage Disclosure}
The authors utilized LLMs solely for the purpose of improving the language fluency and grammar of portions of this manuscript. All ideas, analyses, and conclusions are the authors’ own. The authors have thoroughly reviewed and edited the AI-suggested text and take full responsibility for the content of this publication.

\bibliographystyle{ACM-Reference-Format}
\balance
\bibliography{sample-base}

\appendix

\end{document}